\documentclass[aps,prl,twocolumn,showpacs,superscriptaddress]{revtex4}
\usepackage{graphicx}
\usepackage{bm}
\usepackage{natbib}

\begin{document}

%\preprint{}
\title{Effect of nematic order on the low-energy spin fluctuations in detwinned BaFe$_{1.935}$Ni$_{0.065}$As$_2$}

\author{Wenliang Zhang}
\affiliation{Beijing National Laboratory for Condensed Matter Physics, Institute of Physics, Chinese Academy of Sciences, Beijing 100190, China}
\author{J. T. Park}
\affiliation{Heinz Maier-Leibnitz Zentrum (MLZ), Technische Universit$\ddot{a}$t M$\ddot{u}$nchen, D-85748 Garching, Germany}
\author{Xingye Lu}
\affiliation{Beijing National Laboratory for Condensed Matter Physics, Institute of Physics, Chinese Academy of Sciences, Beijing 100190, China}
\author{Yuan Wei}
\affiliation{Beijing National Laboratory for Condensed Matter Physics, Institute of Physics, Chinese Academy of Sciences, Beijing 100190, China}
\author{Xiaoyan Ma}
\affiliation{Beijing National Laboratory for Condensed Matter Physics, Institute of Physics, Chinese Academy of Sciences, Beijing 100190, China}
\author{Lijie Hao}
\affiliation{China Institute of Atomic Energy, Beijing 102413, China}
\author{Pengcheng Dai}
\affiliation{Department of Physics and Astronomy, Rice University, Houston, Texas 77005-1827, USA}
\author{Zi Yang Meng}
\affiliation{Beijing National Laboratory for Condensed Matter Physics, Institute of Physics, Chinese Academy of Sciences, Beijing 100190, China}
\author{Yi-feng Yang}
\affiliation{Beijing National Laboratory for Condensed Matter Physics, Institute of Physics, Chinese Academy of Sciences, Beijing 100190, China}
\affiliation{Collaborative Innovation Center of Quantum Matter, Beijing, China}
\affiliation{School of Physical Sciences, University of Chinese Academy of Sciences, Beijing 100190, China}
\author{Huiqian Luo}
\email{hqluo@iphy.ac.cn}
\affiliation{Beijing National Laboratory for Condensed Matter Physics, Institute of Physics, Chinese Academy of Sciences, Beijing 100190, China}
\author{Shiliang Li}
\email{slli@iphy.ac.cn}
\affiliation{Beijing National Laboratory for Condensed Matter Physics, Institute of Physics, Chinese Academy of Sciences, Beijing 100190, China}
\affiliation{Collaborative Innovation Center of Quantum Matter, Beijing, China}
\affiliation{School of Physical Sciences, University of Chinese Academy of Sciences, Beijing 100190, China}
\begin{abstract}
The origin of nematic order remains one of the major debates in iron-based superconductors. In theories based on spin nematicity, one major prediction is that the spin-spin correlation length at (0,$\pi$) should decrease with decreasing temperature below the structural transition temperature $T_s$. Here we report inelastic neutron scattering studies on the low-energy spin fluctuations in BaFe$_{1.935}$Ni$_{0.065}$As$_2$ under uniaxial pressure. Both intensity and spin-spin correlation start to show anisotropic behavior at high temperature, while the reduction of the spin-spin correlation length at (0,$\pi$) happens just below $T_s$, suggesting strong effect of nematic order on low-energy spin fluctuations. Our results favor the idea that treats the spin degree of freedom as the driving force of the electronic nematic order.

\end{abstract}



\maketitle

The parent compounds of most iron-based superconductors exhibit long-range antiferromagnetic (AF) order at low temperature with a stripe-type in-plane structure, where the adjacent magnetic moments are anti-parallel and parallel to each other along the orthorhombic a and b axes, respectively \cite{DaiP2015}. In addition to the breaking of the translational symmetry, this configuration also breaks the fourfold rotational symmetry of the underlying lattice. It is suggested theoretically that spin correlations in iron pnictides may form a nematic order by restoring the $O$(3) spin-rotational symmetry while keeping the $C_4$ tetragonal symmetry broken within a narrow temperature range, $T_N \leq T \leq T_s$, where $T_N$ and $T_s$ are the AF and structural transition temperatures, respectively \cite{FangC2008,XuC2008,FernandesRM2011,Fernandes2012-1,Fernandes2012-2,Liangs2013,Fernandes2014}. While inelastic neutron scattering (INS) experiments found clear evidence of anisotropic spin excitations in the unaixial strained paramagnetic orthorhombic phase of iron pnictides \cite{Luxy2014,Songy2015}, the effect is also present in the paramagnetic tetragonal phase above $T_s$ due to the presence of uniaxial strain \cite{Luxy2016}. In addition, neutron scattering and NMR measurements have also found spin excitation anisotropy without uniaxial strain below $T_s$, consistent with the theoretical predictions \cite{Zhangq2015,FuM2012,DioguardiAP2016}.

While there is no question of the presence of spin nematicity in iron pnictides, whether it is the driving force of the electronic nematicity is still under debate \cite{ChuJH2012, KuoHH2016, Fernandes2010,Yoshizawa2012,Kasahara2012}. An alternative picture is to treat the orbital degree of freedom as the primary origin of the electronic nematicity \cite{LvW2009,LeeCC2009}. In angle-resolved photoemission spectroscopy (ARPES) measurements, a pronounced energy splitting of bands with $d_{xz}$ and $d_{yz}$ orbital characters has been detected \cite{YiM2011}. Moreover, Raman and electron diffraction measurements also reveal orbital quadrupole fluctuations in normal and superconducting states \cite{MaCh2014, Thorsmolle2016}. While the nematic order and its fluctuations have been suggested to be important for the mechanism of novel superconductivity \cite{LedererS15,MetlitskiMA15}, the origin of the electronic nematic order is still one of the central unsettled issues in iron-based superconductors \cite{Fernandes2014,BaekSH2015,WangQ2016}.

The difficulty lies in the fact that the spin and orbital degrees of freedom are generally coupled even without the presence of long-range AF order such as the FeSe system \cite{ChubukovAV2015,YamakawaY16}. It is thus important to compare the experimental observation of nematic order with theoretical results. In a stripe-type AF phase, the low-energy spin waves can only be found around ($\pi$,0), whereas equal intensity should be observed at ($\pi$,0) and (0,$\pi$) above $T_N$. One of the most important predictions of spin nematic theory is that the nematic order should enhance magnetic excitations at ($\pi$, 0) in the form of increasing both the intensity and the correlation length while those around (0, $\pi$) are suppressed in an opposite way just below $T_s$ \cite{FernandesRM2011,Fernandes2012-2}. Experimentally, there is a lack of study on the spin-spin correlation at (0,$\pi$), which is crucial to establish the nematic nature of the spin system.

Here we report INS study on the spin nematicity in detwinned BaFe$_{1.935}$Ni$_{0.065}$As$_2$. The difference between ($\pi$,0) and (0,$\pi$) both in the intensity and the correlation length starts well above $T_s$, indicating a possible stabilization of nematic spin fluctuations by the uniaxial pressure due to spin-lattice coupling. The spin-spin correlation at (0,$\pi$) starts to decrease just below $T_s$, suggesting a strong influence of nematic order on low-energy spin fluctuations. Our results are consistent with the spin nematic theories \cite{FernandesRM2011,Fernandes2012-2}.

\begin{figure}
\includegraphics[scale=1.2]{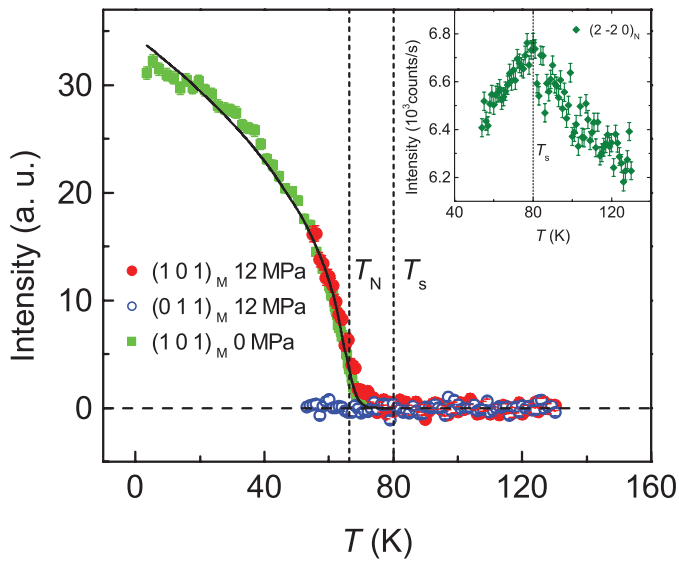}
\caption{Temperature dependence of the intensity of the AF Bragg peaks at (1 0 1) and (0 1 1) under 0 and 12 MPa. The solid line is fitted as described in the text. $T_N$ is determined from the fitting. The inset shows the temperature dependence of the intensity of the nuclear Bragg peak at (2 -2 0), which shows a kink at $T_s$.}
\label{fig1}
\end{figure}

Single crystal of BaFe$_{1.935}$Ni$_{0.065}$As$_2$  was grown by self-flux method as described previously \cite{Chenyc2011}. In this paper, we will always use the orthorhombic notation, in which the momentum transfer $\bm{Q}$ in reciprocal space is defined as $\mathbf{Q}$=\emph{H}$\bm{a^*}$+\emph{K}$\bm{b^*}$+\emph{L}$\bm{c^*}$, where \emph{H}, \emph{K}, \emph{L} are Miller indices and a$^*$=2$\pi$/a, b$^*$=2$\pi$/b, c$^*$=2$\pi$/c with a$\approx$b$\approx$5.54 {\rm\AA} and c=12.3 {\rm\AA}. The slight difference between a and b in the orthorhombic phase has no impact in our measurements. The onset of nematic order is accompanied by a tetragonal-to-orthorhombic structural transition that results in twinning of the crystals. To resolve the spin excitations from ($\pi$, 0) and (0, $\pi$), a uniaxial pressure along one axis of the orthorhombic lattice has to be applied to detwin the sample. Therefore, the sample was cut into rectangular shape along a/b directions of the orthorhombic cell by high precision wire saw, and then loaded into an aluminum device with a spring to apply a uniaxial pressure of 12 MPa\cite{Luxy2014}. The device is mounted on a supporting sample holder to align the crystal in the scattering plane spanned by the wave vector (1 0 1) and (0 1 1), where the spin excitations at $\mathbf{Q}$=(1 0 1) and (0 1 1) can be measured within one scattering plane \cite{Luxy2014}. Neglecting the dependence along (0 0 L), these two wavevectors correspond to ($\pi$,0) and (0,$\pi$) as discussed above, respectively. The INS experiments are carried out at PUMA thermal triple-axis spectrometer at MLZ \cite{SobolevO2015}. All measurements were done with a fixed final wave vector, $k_f$=2.662{\rm\AA}, and horizontally and vertically curved pyrolytic graphite (PG) crystals were used as monochromator and analyzer.

\begin{figure}
\includegraphics[scale=.6]{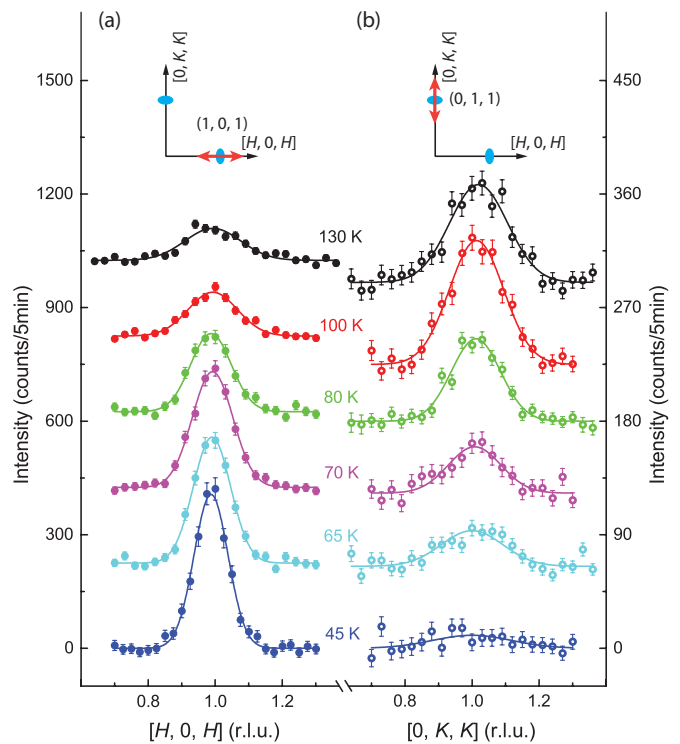}
\caption{Background-subtracted $\mathbf{Q}$-scans at 7 meV along (a) (H 0 H) and (b) (0 K K). The data are shifted for different temperatures. The solid lines are the Gaussian fits of the data.}
\label{fig2}
\end{figure}

Figure 1 gives the temperature dependence of the intensity of magnetic Bragg peaks at (1 0 1) and (0 1 1) at both zero and 12 MPa. Under zero pressure, the twinning of the crystals results in the same intensity at (1 0 1) and (0 1 1) (not shown). The zero intensity of ( 0 1 1) under 12 MPa at all temperatures suggests that the sample is fully detwinned. The intensity at (1 0 1) is proportional to $M^2$, where $M$ is the AF order parameter. Therefore, it may be fitted as (1-$T$/$T_N$)$^{2\beta_{AF}}$ with a Gaussian distribution of $T_N$ \cite{BirgeneauR1973,DhitalC2014}. The mean value of $T_N$ (66.4 K), the critical exponent $\beta_{AF}$ (0.22) and the Gaussian width $\sigma$ (2.6 K) are consistent with previous reports in the Ba(Fe$_{1-x}$Co$_x$)$_2$As$_2$ system \cite{DhitalC2014}. The fitted $T_N$ is also close to that obtained in the resistivity measurement \cite{Chenyc2011}. The pressure of 12 MPa causes slight enhancement of the intensity at (1 0 1) above $T_N$ but the data below $T_N$ between 0 and 12 MPa are the same after proper scaling. Therefore, $\beta$ should not change significantly with pressure \cite{DhitalC2014}. Although the applied uniaxial pressure should render the tetragonal-to-orthorhombic structural transition to a crossover \cite{Luxy2016}, we can still observe a strong extinct effect for the intensity of the nuclear Bragg peak (2 -2 0) as shown in the inset of Fig. 1, suggesting a further orthorhombic structural distortion below this temperature, which is labeled as $T_s$.

\begin{figure}
\includegraphics[scale=0.8]{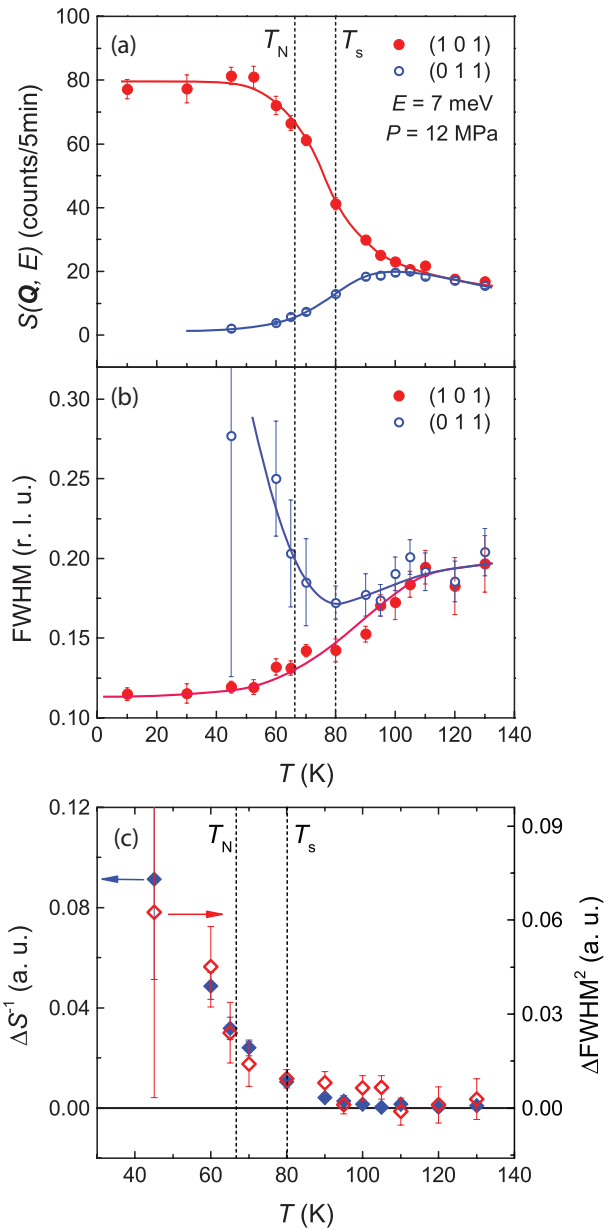}
\caption{Temperature dependence of (a) S($\mathbf{Q}$, E) and (b) FWHM at (1 0 1) and (0 1 1) at 7 meV and 12 MPa. (c) Temperature dependence of $\Delta$S$^{-1}$ (blue solid diamonds) and $\Delta$FWHM$^2$ (red open diamonds) as defined in the main text. The solid line are guided to the eye. The vertical dashed lines indicate $T_s$ and $T_N$.
}
\label{fig3}
\end{figure}

Figure 2 shows the $\mathbf{Q}$-scans around (1 0 1) and (0 1 1) at 7 meV. Both the intensity and the width of the peaks show strong temperature dependence. Limited by the scattering plane, the $\mathbf{Q}$ scans can only be done along (H 0 H) and (0 K K) direction for (1 0 1) and (0 1 1), respectively. Nevertheless, since spin correlation along c direction is much weaker than that in the a-b plane and almost no correlation along c axis is found above $T_N$ \cite{MatanK2009}, the $\mathbf{Q}$ scans along (H 0 H) and (0 K K) mainly reflect the behaviors of S($\mathbf{Q}$, E) along the (H 0) and (0 K) directions, respectively. In other words, the width of the peaks shown in Fig. 2 gives a reasonable measure of the in-plane spin correlations at ($\pi$, 0) and (0, $\pi$).

Figure 3(a) and 3(b) show the temperature dependence of peak intensity S($\mathbf{Q}$,E) and the fitted full width at half maximum (FWHM) at (1 0 1) and (0 1 1). The difference of S($\mathbf{Q}$,E) between two wavevectors becomes larger with decreasing temperature below about 110 K that is much higher than $T_s$. The FWHM at (1 0 1) smoothly decreases with decreasing temperature, showing no sign of either $T_N$ or $T_s$. On the other hand, similar measurements in LaFeAsO and Ba(Fe$_{0.953}$Co$_{0.047}$)$_2$As$_2$ show dramatic decrease of line width only between $T_s$ and $T_N$ \cite{Zhangq2015}. The difference between these two results is due to fact that the measurements in this work are done under 12 MPa whereas no pressure is applied in the latter. As reported previously, a uniaxial pressure smears out the structural transition and induces an orthorhombic lattice distortion at all temperatures \cite{Luxy2016}. While it seems that the nematic transition is affected in a similar way under uniaxial pressure by checking the temperature dependence of FWHM$_{(101)}$, the FWHM at (0 1 1) clearly shows a sudden rise just below $T_s$ although the difference between them can be already observed below about 110 K, which clearly suggests the establishment of nematic order with the suppression of spin-spin correlation at (0 1 1). Therefore, the nematic order transition seems still well defined even under large uniaxial pressure.

The nematic nature of spin-spin correlation may be further revealed by comparing the temperature dependence of $\Delta$S($\mathbf{Q}$,E)$^{-1}$ and $\Delta$FWHM$^2$, where $\Delta$S$^{-1}$ = S($\mathbf{Q}$,E)$_{(011)}$$^{-1}$ - S($\mathbf{Q}$,E)$_{(101)}$$^{-1}$and $\Delta$FWHM$^2$ = FWHM$_{(011)}^2$-FWHM$_{(101)}^2$. It has been suggested that for overdamped spin excitations, S($\mathbf{Q}$,E) $\propto (r\pm\varphi)^{-1}$ and FWHM $\propto (r\pm\varphi)^{1/2}$, where $\varphi$ and $r$ denote the nematic order parameter and the magnetic correlation length with $\varphi$=0, respectively \cite{Fernandes2012-2}. This leads to a relationship of $\Delta$S$^{-1} \propto \Delta$FWHM$^2$, as shown in Fig. 3(c).

The nematic order parameter for the spin system can be expressed as $\varphi \propto M_1^2-M_2^2$, where $M_1^2$ and $M_2^2$ are the spin fluctuations at ($\pi$,0) and (0,$\pi$) respectively \cite{FernandesRM2011,Fernandes2012-2}. Therefore, we define $\chi"_{nem}$ = $\chi"_{(101)}$-$\chi"_{(011)}$ to approximately represent $\varphi$, where $\chi"$ at (1 0 1) and (0 1 1) are obtained by integrating the $\mathbf{Q}$-scans in Fig. 2 corrected by the Bose factor. While the nematic transition happens at about 80 K, the non-zero nematic order parameter can be observed at much higher temperature as shown in Fig. 4 \cite{Luxy2014}, which is most likely due to the pressure as described by the Landau theory of phase transition with an external field \cite{Luxy2016,Renx2015}. The free energy may be simply given as $F$ = a$\varphi^2$+b$\varphi^4$+$h\varphi$, where $h$ denotes the conjugated field. Here a=a$_0$($T$-$T_s$) and b are the parameters as in a conventional Landau theory. The solid line in Fig. 4 suggests that this simple model can indeed describe the nematic behavior above $T_s$, demonstrating the role of uniaxial stress as the external field for the nematic order parameter \cite{Fernandes2012-2}.

\begin{figure}
\includegraphics[scale=1.1]{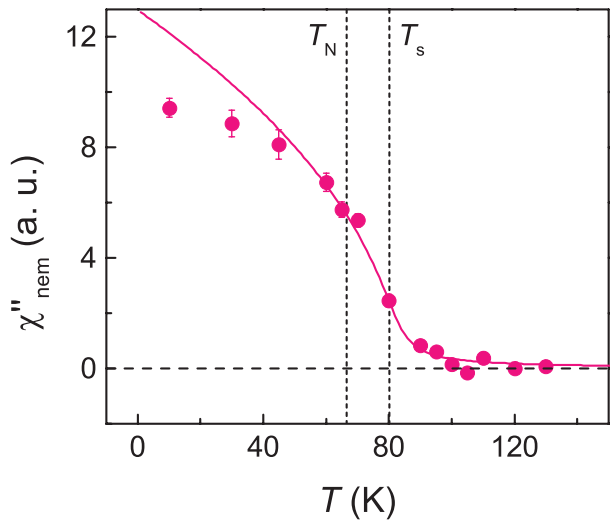}
\caption{ Temperature dependence of $\chi"_{nem}$. The solid line is calculated according to the Landau free energy as described in the main text with a$_0$ = 3.73, b = 0.036 and $h$ = 266. The vertical dashed lines indicate $T_s$ and $T_N$.
}
\label{fig4}
\end{figure}

In the spin nematic picture, both the intensity and the spin-spin correlation length should exhibit anisotropy between (1 0 1) and (0 1 1) wavevectors in the nematic phase \cite{FernandesRM2011,Fernandes2012-2}. INS studies on detwinned BaFe$_{2-x}$Ni$_x$As$_2$ show that the intensity of low-energy spin excitations at ($\pi$,0) is indeed larger than that at (0,$\pi$) below a temperature far above $T_s$, clearly demonstrating the change from fourfold to twofold symmetry of the spin system under uniaxial pressure \cite{Luxy2014}. The enhancement of the spin-spin correlation length at ($\pi$,0) is observed between $T_s$ and $T_N$ in LaFeAsO and Ba(Fe$_{0.953}$Co$_{0.047}$)$_2$As$_2$ at zero pressure \cite{Zhangq2015}. Generally speaking, the enhancement of spin-spin correlation could happen at temperatures above $T_N$ if another long-range order that is strongly coupled to the spin system is established. The findings of the reduction of spin-spin correlation at (0 1 1) blow $T_s$ and the correlation between $\Delta$S$^{-1}$ and $\Delta$FWHM$^2$ are consistent with the spin nematic theories. This suggests that the nematic order enhances the spin fluctuations at (1 0 1) with increasing spin-spin correlation while suppressing those at (0 1 1) with decreasing spin-spin correlation. Therefore, we have clearly demonstrated the nematic nature of the spin-spin correlation in our sample.

In conclusion, the most important result of this paper is the observation of the decease of the low-energy spin-spin correlation length at (0,$\pi$) below $T_s$ as predicted by the spin nematic theory. Combining with previous results \cite{Luxy2014,Zhangq2015}, the predictions of spin dynamics in the nematic phase by the spin nematic theory \cite{FernandesRM2011,Fernandes2012-2} have been experimentally confirmed. Our results favor the idea that the electronic nematic order is driven by the spin degree of freedom.

\acknowledgements

This work is supported by the "Strategic Priority Research Program (B)" of Chinese Academy of Sciences (XDB07020300), Ministry of Science and Technology of China (2012CB821400, 2011CBA00110,2015CB921302,2016YFA0300502), National Science Foundation of China (No. 11374011, 11374346, 91221303,11421092,11574359), and the National Thousand-Young-Talents Program of China. The work at Rice University is supported by the
U.S. NSF-DMR-1362219, DMR-1436006, and in part by the Robert A. Welch Foundation Grant Nos. C-1839 (P.D.).

\bibliography{spinnematic}

\end{document}